\documentclass[12pt]{article}
\usepackage{amssymb,cite}
\def\rme{{\rm e}}
\def\rmi{i} 
\def\rmd{{\rm d}}
\newsavebox{\uuunit}
\sbox{\uuunit}
    {\setlength{\unitlength}{0.825em}
     \begin{picture}(0.6,0.7)
        \thinlines
        \put(0,0){\line(1,0){0.5}}
        \put(0.15,0){\line(0,1){0.7}}
        \put(0.35,0){\line(0,1){0.8}}
       \multiput(0.3,0.8)(-0.04,-0.02){12}{\rule{0.5pt}{0.5pt}}
     \end {picture}}
\newcommand {\unity}{\mathord{\!\usebox{\uuunit}}}

\csname @addtoreset\endcsname{equation}{section}
\newif\ifpdf
\ifx\pdfoutput\undefined
   \pdffalse
 \else
   \pdfoutput=1
   \pdftrue
  \usepackage[pdftex]{hyperref}
\fi

\newcommand{\bref}[1]{(\ref{#1})}
\def\e{\varepsilon}
\def\s{\sigma}
\def\a{\alpha}
\def\d{\delta}
\def\pa{\partial}
\def\w{\omega}
\newcommand{\nn}{\nonumber}

\def\t{\theta}
\def\bt{\bar{\theta}}
\def\dt{{\rm d}\theta}
\def\dbt{{\rm d}\bar{\theta}}

\def\cF{\mathcal{F}}
\def\cT{\mathcal{T}}
\def\hg{\hat{g}}
\def\hga{\hat{\gamma}}
\def\vu{\vec{u}}
\def\cL{\mathcal{L}}
\newcommand{\ep}{\epsilon}
\begin{document}

\begin{titlepage}
\begin{flushright}
UB-ECM-PF-05/18\\
KUL-TF-05/17\\
hep-th/0507135
\end{flushright}
\vspace{.5cm}
\begin{center}
\baselineskip=16pt
{\LARGE Non Relativistic D$p$ Branes 
}\\
\vfill
{\large  Joaquim Gomis $^{1,2}$, 
Filippo Passerini $^2$, Toni Ramirez$^1$,\\
 Antoine Van Proeyen$^2$
  } \\
\vfill
{\small $^1$  Departament ECM, Facultat F{\'\i}sica,
  Universitat de Barcelona \\
  and\\
  CER for Astrophysics, Particle Physics and Cosmology, ICE/CSIC
\\
Diagonal 647, E-08028
\\ \vspace{6pt}
$^2$ Instituut voor Theoretische Fysica, Katholieke Universiteit Leuven,\\
       Celestijnenlaan 200D B-3001 Leuven, Belgium.
      \\[2mm] }
\end{center}
\vfill
\begin{center}
{\bf Abstract}
\end{center}
{\small We construct a kappa-symmetric and diffeomorphism-invariant
non-relativistic D$p$-brane action as a non-relativistic limit of a
relativistic D$p$-brane action in flat space. In a suitable gauge the
world-volume theory is given by a supersymmetric free field theory in
flat spacetime in $p+1$ dimensions of bosons, fermions and gauge fields.
 }\vspace{2mm} \vfill \hrule width 3.cm
{\footnotesize \noindent e-mails: gomis@ecm.ub.es,
filippo.passerini@fys.kuleuven.be, tonir@ecm.ub.es,\\
antoine.vanproeyen@fys.kuleuven.be }
\end{titlepage}
\addtocounter{page}{1}
 \tableofcontents{}
\newpage
\section{Introduction}

Non-relativistic string theory \cite{Gomis:2000bd,Danielsson:2000gi} is a
consistent sector of string theory, whose world-sheet conformal field
theory description has the appropriate Galilean symmetry
\cite{Brugues:2004an}. Non-relativistic superstrings and non-relativistic
superbranes \cite{Gomis:2004pw,Garcia:2002fa} are obtained as a certain
decoupling limit of the full relativistic theory. The basic idea behind
the decoupling limit is to take a particular non-relativistic limit in
such a way that the light states satisfy a Galilean-invariant dispersion
relation, while the rest decouple. For the case of strings, this can be
accomplished by considering wound strings in the presence of a background
$B$-field and tuning the $B$-field so that the energy coming from the
$B$-field cancels the tension of the string. In flat space,  once kappa
symmetry and diffeomorphism invariance are fixed, non-relativistic
strings are described by a free field theory in flat space. In
AdS$_5\times S^5$ \cite{Gomis:2005pg}, the world-sheet theory reduces to
a supersymmetric free field theory in AdS$_2$.

In this paper we study the non-relativistic limit of non-perturbative
supersymmetric objects of string theory. We  study non-relativistic
supersymmetric D$p$ branes in flat spacetime. The point of departure  is
to consider the world-volume kappa invariant action of a relativistic
D$p$ brane in flat spacetime
\cite{Aganagic:1996pe,Aganagic:1996nn,Cederwall:1996ri,Bergshoeff:1996tu}.
Since the D$p$ branes are charged under the RR forms, we also consider
its coupling to a closed $p+1$ RR form, $C_{p+1}$. In this way we can
find a limit where the tension of the wound D$p$ brane is cancelled by
the coupling to the $C_{p+1}$ field. Only states with positive charge
remain light in the limit, while the non-positively charged states become
heavy. We obtain a world-volume kappa symmetric action of a
non-relativistic D$p$ brane. When  kappa symmetry \cite{Siegel:1983hh}
and diffeomorphisms are fixed, the non-relativistic D$p$-brane action is
described by a supersymmetric {\it free} field theory in flat spacetime
in $p+1$ dimensions of bosons, fermions and gauge fields. This is the
main result.

The paper is organized as follows. In section \ref{ss:relDpbrane}, we
summarize the basic properties of kappa-symmetric relativistic D$p$-brane
actions in flat space. In section \ref{ss:NrelDpbranes}, we consider the
non-relativistic limit of relativistic D$p$ branes. The supersymmetry and
kappa transformations are discussed in section \ref{ss:susykappa}. It is
shown there how after the gauge fixing these transformations give rise to
a rigid supersymmetric vector multiplet with the usual supersymmetry
algebra. In section \ref{ss:D1 string}, we will specify to the case of a
D string. While in other cases the Wess-Zumino (WZ) term is given
implicitly as a $(p+2)$-form over an embedding manifold, in this case the
form of the WZ term is simple and we give it explicitly. We finish by
some conclusions and an appendix with conventions.
\medskip
\section{ Relativistic D$p$ branes}
\label{ss:relDpbrane}


The action for a relativistic D$p$ brane propagating in flat
space\footnote{ We are using conventions close to these of
\cite{Aganagic:1996pe,Aganagic:1996nn}, except that the exterior
derivative commutes with $\theta$, and our spinor conventions imply that
their $\bar \theta $ is $-\rmi\bar \theta $ for us. See the appendix for
more details.} is
\cite{Aganagic:1996pe,Aganagic:1996nn,Cederwall:1996ri,Bergshoeff:1996tu}
\begin{eqnarray}
S&=&-T_{p}\int \rmd^{p+1} \s \sqrt{-\det{(G_{ij}+\cF_{ij})}} + T_{p}\int
\Omega_{p+1}\nonumber\\
 &=&-T_{p}\int\left[ \cL_{\mathrm{NG}}-\cL_{\mathrm{WZ}}
\right]=-T_{p}\int\cL_{\mathrm{GS}}.\label{Dpaction}
\end{eqnarray} $G_{ij}$ is the induced
metric constructed from the supertranslation invariant 1-form
\begin{equation}\label{pi1form}
{\Pi}^m={\rm d}X^m+i\bt\Gamma^m\rmd\theta.
\end{equation}
$\cF_{ij}$ is constructed from the two form $\cF=2\pi\a'F-b$, which is
written in terms of the field strength of the Born-Infeld (BI) field,
$A$, and of the pullback of the fermionic components of the $B$ field in
superspace. $\cL_{\mathrm{WZ}}=\Omega_{p+1}$ is  the WZ term. Since the
expression is complicated\footnote{The explicit form of the WZ term is
given in \cite{Hatsuda:1998by,Kamimura:1997ju}.}, it is useful to
introduce a $(p+2)$-form $h_{p+2}$ such that
\begin{equation}
h_{p+2}=\rmd \Omega_{p+1}.
\end{equation}

For type IIA D$p$ branes ($p$ even), the  forms are given by
\begin{eqnarray}
b&=&-i\bt\Gamma_{11}\Gamma_m \rmd\theta
\left(\Pi^m-\frac{i}2\bt\Gamma^m \rmd\theta\right),\nonumber\\
 h_{p+2}&=&(-)^n
\rmi\rmd\bt\cT_p\rmd\theta, \qquad\ p=2n,
\end{eqnarray}
where $\cT_p$ is a $p$ form. To define it, we introduce the formal sum of
differential forms
\begin{eqnarray} \cT_A =\sum_{p=\mathrm{even}}\cT_p=\rme^{\cF}C_A,
\end{eqnarray}
where
\begin{equation}
C_A=\Gamma_{11}+\frac{1}{2!}\psi^2+\frac{1}{4!}\Gamma_{11}\psi^4+\frac{1}{6!}\psi^6+\ldots.
\end{equation}
and
\begin{equation}
\psi=\Pi^m \Gamma_m. \label{defpsi}
\end{equation}
The  D$p$-brane  action \bref{Dpaction} is invariant under the
supersymmetry transformations
\begin{eqnarray}\nn\label{susyrIIA}
\delta_\ep\t&=&\ep, \qquad  \delta_\ep X^m=-i\bar{\ep}\Gamma^m\t,\\
\delta_\ep (2\pi\a'A)&=& -i \bar{\ep}\Gamma_{11}\Gamma_m\t\rmd X^m
+\frac{1}{6}\left(\bar{\ep}\Gamma_{11}\Gamma_m\t\bt\Gamma^m\rmd \t+
\bar{\ep}\Gamma_m\t\bt\Gamma_{11}\Gamma^m\rmd \t\right),
\end{eqnarray}
and under the kappa transformations
\begin{eqnarray}
\delta_\kappa\bt&=&\frac12\bar{\kappa}\left[1+(-)^n\Gamma_\kappa\right],
\qquad  \d_\kappa  X^m=-i\bt\Gamma^m \delta_\kappa\t, \nn \\
\delta_\kappa (2\pi\a'A)&=&+i\delta_\kappa\bt\Gamma_{11}\Gamma_m\t\Pi^m
+\frac{1}{2}\delta_\kappa\bt\Gamma_{11}\Gamma_m\t\bt\Gamma^m\rmd \t-
\frac{1}{2}\delta_\kappa\bt\Gamma_m\t\bt\Gamma_{11}\Gamma^m\rmd
\t,\label{krelIIA}
\end{eqnarray}
where
\begin{eqnarray}\label{gk}
\Gamma_\kappa&=&\frac{1}{(p+1)!}\frac{\varepsilon^{i_0\ldots
i_p}}{\sqrt{-\det(G+\cF)}}\left( \rho_{p+1}\right) _{i_0\ldots i_p}.
\end{eqnarray}
The $(p+1)$-form $\rho_{p+1}$ is defined
\cite{Aganagic:1996pe,Aganagic:1996nn} by the formal sum
\begin{equation}
\rho_A =\sum_{p=\mathrm{even}}\rho_{p+1}=\rme^{\cF}S_A,
\end{equation}
where
\begin{equation}
S_A=\Gamma_{11}\psi+\frac{1}{3!}\psi^3+\frac{1}{5!}\Gamma_{11}\psi^5+\frac{1}{7!}\psi^7+\ldots.
\end{equation}

For the type IIB D$p$ branes ($p$ odd) we have
\begin{eqnarray}
b&=&-i\bt\Gamma_m \tau_3\rmd\theta\left(\Pi^m-\frac{i}2\bt\Gamma^m
\rmd\theta\right), \nonumber\\
h_{p+2}&=&i\dbt \cT_p \rmd\theta,
\end{eqnarray}
where
\begin{equation}
\cT_B=\sum_{p=\mathrm{odd}}\cT_p=\rme^{\cF}S_B\tau _1,
\end{equation}
and
\begin{equation}
S_B(\psi)=\psi+\frac1{3!}\tau_3\psi^3+\frac1{5!}\psi^5+\frac1{7!}\tau_3\psi^7+\dots,
\end{equation}
where $\psi$ is defined in (\ref{defpsi}). The supersymmetry
transformations are given by
\begin{eqnarray}\nn\label{susyrIIB}
\delta_\ep\t&=&\ep, \qquad  \delta_\ep X^m=-i\bar{\ep}\Gamma^m\t,\\
\delta_\ep (2\pi\a'A)&=& -i \bar{\ep}\tau_3\Gamma^m\t\rmd X^m
+\frac{1}{6}\left(\bar{\ep}\tau_3\Gamma_m\t\bt\Gamma^m\rmd \t+
\bar{\ep}\Gamma_m\t\bt\tau_3\Gamma^m\rmd \t\right).
\end{eqnarray}
The kappa   transformations are
\begin{eqnarray}
\delta_\kappa\bt&=&\frac12\bar{\kappa}\left(1+\Gamma_\kappa\right),\qquad
\d_\kappa  X^m=-i\bt\Gamma^m \delta_\kappa\t ,\nn \\
\delta_\kappa (2\pi\a'A)&=&i\delta_\kappa\bt\tau_3\Gamma_m\t\Pi^m
+\frac{1}{2}\delta_\kappa\bt\tau_3\Gamma_m\t\bt\Gamma^m\rmd \t-
\frac{1}{2}\delta_\kappa\bt\Gamma_m\t\bt\tau_3\Gamma^m \rmd
\t,\label{krelIIB}
\end{eqnarray}
where
\begin{equation}
\Gamma_\kappa= \frac{1}{(p+1)!}\frac {\varepsilon^{i_0\ldots
i_p}}{\sqrt{-\det{\left(G+\cF\right)}}} \left( \rho_{p+1}\right)
_{i_0\ldots i_p},
\end{equation}
and $\rho_{p+1}$ is defined as a $(p+1)$-form given by
\begin{equation}
\rho_B=\sum_{p=\mathrm{odd}}\rho_{p+1}=\rme^{\cF}C_B(\psi)\tau_1,
\end{equation}
where $C_B$ is
\begin{equation}
C_B(\psi)=\tau_3+\frac1{2!}\psi^2+\frac1{4!}\tau_3\psi^4+\frac1{6!}\psi^6+
\dots.
\end{equation}

We can switch on one more coupling in the world-volume consistent with
all the symmetries of the  D$p$-brane action. From the spacetime point of
view, it corresponds to turning on a closed $(p+1)$ RR field, which does
not modify the flat supergravity equations of motion.
\begin{equation}
  \cL=-T_p \left[ \cL_{\mathrm{NG}}-\cL_{\mathrm{WZ}} -\cL_{C_{p+1}}\right],
\end{equation}
where $\cL_{C_{p+1}}=f^*C_{p+1}$ is the pullback of $C_{p+1}$ on the
world-volume (for more details see below).

\section{Non-relativistic D$p$ branes}\label{ss:NrelDpbranes}

In this section we derive the action for non-relativistic D$p$ branes.
The non-relativistic limit of
strings\cite{Gomis:2000bd,Danielsson:2000gi,Gomis:2004pw} is obtained by
decoupling some charged light degrees of freedom that obey a
non-relativistic dispersion relation from the full relativistic theory.
This is achieved by rescaling the world-volume fields with a
dimensionless parameter $\omega $ and later sending the parameter to
infinity. This limit implies that the transverse oscillations are small.
For the case of D$p$ branes we should do the following rescaling
\begin{eqnarray}\label{scal}
X^{\mu}&=& \omega  x^{\mu}, \nn \\
X^a &=& X^a, \nn \\
T_p &=& \omega^{1-p}T_{\mathrm{NR}},\\
\nn (2\pi \a') F_{ij}&=& \w f_{ij},\\
(2\pi \alpha ') A_i&=& \omega W_i
\nonumber\\
\theta&=&\sqrt{\omega}\theta_- +
\frac{1}{\sqrt{\omega}}\theta_+,\nn\\
C_{\mu_0\cdots\mu_p}&=&-\varepsilon_{\mu_0\cdots\mu_p},\nonumber
\end{eqnarray}
where $X^m$ has been split in $X^\mu$ and $X^a$. The $X^\mu $ are the
coordinates of target space parallel to the brane and $X^a$ are the
transverse coordinates. The NR gauge field strength is $f_{ij}=\partial
_iW_j-\partial _jW_i$. The scaling of the fermions depends on the
splitting of the fermions due the matrix $\Gamma_*$:
\begin{equation}
  \Gamma_\ast\t_\pm =\pm\t_\pm.
 \label{Gammastartheta}
\end{equation}
The expression for $\Gamma_\ast$ is
\begin{equation}\label{gammaIIA}
\Gamma_\ast=(-)^{n+1}\Gamma_{0\ldots p}\Gamma_{11}^{n+1},\qquad p=2n,
\end{equation}
for  type IIA D$p$ branes and
\begin{equation}\label{gammaIIB}
\Gamma_\ast=\Gamma_{0\ldots p}i\tau_{3}^{n}\tau_{2},\qquad p=2n-1,
\end{equation}
for type IIB D$p$ branes. $\tau _{1,2,3}$ are the Pauli matrices.
$\Gamma_\ast$ appears as the first term of the non-relativistic expansion
of the matrix $\Gamma _\kappa $ appearing in the kappa transformations as
will be shown below. Properties of the projected spinors are given in the
appendix.

In order to compute the non-relativistic limit we should see how the
forms involved in the action rescale under \bref{scal}. The
supertranslation 1-form \bref{pi1form}  scales as
\begin{equation}
\Pi^\mu =\omega \hat{e}^\mu+\frac {i}{\omega}\bt_+\Gamma^\mu
\rmd\theta_+,\quad \Pi^b = u^b,
\end{equation}
where we have introduced
\begin{eqnarray}
\hat{e}^{\mu}&= &e^{\mu}+i\,\bar \t_-\Gamma^\mu \rmd\theta_-,\qquad
e^\mu = \rmd x^\mu, \nn\\
u^a&=&\rmd x^a+2i\bt_+\Gamma^a \rmd \t_-,\qquad
x^a=X^a+i\,\bt_-\Gamma^a\theta _+.
\end{eqnarray}
The form $\cF$ scales as
\begin{equation}\label{f}
\cF=\omega\cF^{(1)}+\frac{1}{\omega}\cF^{(-1)},
\end{equation}
where for IIA
\begin{eqnarray}\nn
\cF^{(1)}&=&f+\left[\left(i\bt_
-\Gamma_\mu\Gamma_{11}\rmd\theta_++i\bt_+\Gamma_\mu\Gamma_{11}\rmd\theta_-\right)
\left(\hat{e}^\mu-\frac{i}{2}\bt_-\Gamma^\mu \rmd\theta_-\right)+\right.\\
&&\left. +i\bt_-\Gamma_a\Gamma_{11}
\rmd\theta_-\left(u^a-\frac{i}{2}\bt_-\Gamma^a
\rmd\theta_+-\frac{i}{2}\bt_+\Gamma^a \rmd\theta_-\right)\right]
\\\nn
\cF^{(-1)}&=& \frac{1}{2}\left(\bt_-\Gamma_\mu
\Gamma_{11}\rmd\theta_++\bt_+\Gamma_\mu\Gamma_{11}\rmd\theta_-\right)
\bt_+\Gamma^\mu
\rmd\theta_++\\
&&+ i\bt_+\Gamma_a\Gamma_{11}
\rmd\theta_+\left(u^a-\frac{i}{2}\bt_-\Gamma^a
\rmd\theta_+-\frac{i}{2}\bt_+\Gamma^a \rmd\theta_-\right).
\end{eqnarray}
In order to have the expressions for IIB, we should replace $\Gamma_{11}$
by $\tau_3$.

Throughout the analysis, we keep $\omega$ large but finite in the
intermediate computations and only send $\omega$ to infinity at the end.
Therefore, we keep explicitly terms in the action that scale as positive
powers of $\omega$ (which look superficially divergent) and terms that
are independent of $\omega$ (which are finite). We drop terms that scale
as inverse powers of $\omega$ because they cannot contribute when taking
the limit at the end of the analysis.

The NG part of the \bref{Dpaction} becomes after the rescalings
\begin{eqnarray}\nn\label{ngexpt}T_p\mathcal{L}_{\mathrm{NG}}
&=&T_p\sqrt{-\det{\left(G_{ij}+\cF_{ij}\right)}}\nn\\
&=&T_{\mathrm{NR}}\w^2 \mathcal{L}_{\mathrm{NG}}^{\mathrm{div}}
+T_{\mathrm{NR}}\mathcal{L}_{\mathrm{NG}}^{\mathrm{fin}}+{\cal O}(\omega
^{-2}).
\end{eqnarray}
The finite contribution is given by
\begin{equation}
 \mathcal{L}_{\mathrm{NG}}^{\mathrm{fin}}=\frac1{2}\hat{e} \hg^{jl}\vu_l\vu_j
 +i\hat{e}\bt_+\hga^k\pa_{k}\t_++\frac1{4}\hat{e}
\cF_{ij}^{(1)}\cF_{k\ell }^{(1)}\hg^{ik}\hg^{j\ell },
\end{equation}
where $\hat{g}_{jk}=\eta_{\mu\nu}\hat{e}^\mu_j\hat{e}^\nu_k$,
$\hat{e}=\det\hat{e}^\mu_j$ and $\hat\gamma_j= \hat{e}^\mu_j\Gamma_\mu$.
We use the vector signs to indicate sums over the transverse space
components. The superficially divergent contribution, written as a form,
is given by
\begin{equation}
 \rmd^{p+1}\sigma{\cal L}^{\mathrm{div}}_{\mathrm{NG}}=\hat{e}^0\cdots
 \hat{e}^p=-\frac{1}{(p+1)!}\varepsilon _{\mu _0\ldots \mu _p}\hat{e}^{\mu_0}
 \cdots \hat{e}^{\mu _p} .
\end{equation}

Now we consider the scaling of the WZ term. For the IIA case we have
\begin{equation}
T_ph_{p+2}=T_{\mathrm{NR}}\w^2h_{p+2}^{(2)}+T_{\mathrm{NR}}h_{p+2}^{(0)}
+{\cal O}(\omega ^{-2}).
\end{equation}
First, we analyse the superficially divergent term.  This term comes from
the expansion of the term in $h_{p+2}$ that contains $\psi$ to the power
$p$.
\begin{equation}
h_{p+2}^{(2)}=  -\rmi\rmd\bt_- \frac{1}{p!}\hat{e}^{\mu_1}\ldots
\hat{e}^{\mu_p} \e_{\mu_1\ldots\mu_p\nu}\Gamma^{\nu}\rmd\theta_-
=-\e_{\nu\mu_1\ldots\mu_p} \frac{1}{p!}\rmd \hat{e}^\nu
\hat{e}^{\mu_1}\ldots \hat{e}^{\mu_p}.\label{h2p}
\end{equation}
We note that
\begin{equation}\label{gs2n}
\rmd(\rmd^{p+1}\sigma\cL_{\mathrm{GS}}^{\mathrm{div}})=
\rmd(\rmd^{p+1}\sigma\cL_{\mathrm{NG}}^{\mathrm{div}})- h_{p+2}^{(2)}=0.
\end{equation}
As the last term involves only terms with fermions, this cancellation
removes the terms with fermions in $\cL_{\mathrm{NG}}^{\mathrm{div}}$.
There remains the purely bosonic term in
$\cL_{\mathrm{NG}}^{\mathrm{div}}$, which is $e^0\cdots e^p$. Therefore,
the potentially divergent term of $\cL_{\mathrm{GS}}^{\mathrm{div}}$ is
$e^0\cdots e^p$, which is a total derivative. This term can be cancelled
by turning on a closed RR $C_{p+1}$ form, given in (\ref{scal}), which
only leads to the following potentially divergent term
\begin{equation}
  \cL_{C_{p+1}}^{\mathrm{div}}=-\frac{1}{(p+1)!}\varepsilon_{\mu_0\cdots\mu_p}e^{\mu_0}_0\cdots
e^{\mu_p}_p. \label{cLCp1}
\end{equation}

Note that all the positively charged states are light. All states with
non-positive charges become infinitely heavy and decouple.

The finite part of the action of a  NR D$p$ brane is
\begin{equation}
S_{\mathrm{NR}}=-T_{\mathrm{NR}}\int
\rmd\sigma^{p+1}\left(i\hat{e}\bt_+\hga^k\pa_{k}\t_++\frac1{2}\hat{e}\hg^{jl}\vu_l\vu_j
+\frac1{4}\hat{e} \cF_{ij}^{(1)}\cF_{k\ell
}^{(1)}\hg^{ik}\hg^{j\ell}\right)+T_{\mathrm{NR}}\int
\Omega_{p+1}^{(0)},\label{NRNGaction}
\end{equation}
where $\Omega_{p+1}^{(0)}$ is the non-relativistic WZ term. It has a
complicated expression that we give below for the case of D1. In general,
it verifies $\rmd\Omega_{p+1}^{(0)}=h_{p+2}^{(0)}$, where
\begin{eqnarray}
\mathrm{type\ IIA}&:&  h_{p+2}^{(0)}= (-)^n i\frac{1}{p!}\rmd\bt_+
\Gamma_{11}^{(n+1)}\hat{e}^{\mu_1}\ldots
\hat{e}^{\mu_p}\Gamma_{\mu_1\ldots\mu_p}\rmd\t_++\dots, \nonumber\\
 \mathrm{type\ IIB}&:&
h^{(0)}_{p+2}=i\frac1{p!}\dbt_+\left(\tau_3\right)^ni\tau_2
\hat{e}^{\mu_1}\dots
 \hat{e}^{\mu_p}\Gamma_{\mu_1\ldots\mu_p}\dt_++\dots,
 \label{IIABhform}
\end{eqnarray}
and the dots indicate terms with dependence on $\theta _-$.

\subsection{ Supersymmetry and kappa transformations}
\label{ss:susykappa}

The relativistic D$p$-brane action \bref{Dpaction} is invariant under the
supersymmetry \bref{susyrIIA} and kappa transformations \bref{krelIIA}
for type IIA and  (\ref{susyrIIB}) and  (\ref{krelIIB}), respectively,
for type IIB. In order to obtain the non-relativistic counterpart of
these transformations that leave the NR D$p$-brane action,
\bref{NRNGaction}, invariant, we should rescale the supersymmetry
parameter
\begin{equation}
\epsilon=\sqrt{\omega}\epsilon_-+\sqrt{\frac 1{\omega}}\epsilon_+,
\end{equation}
and the kappa parameter
\begin{equation}
\kappa=\sqrt{\omega}\kappa_-+\sqrt{\frac 1{\omega}}\kappa_+.
\end{equation}
We also need the expansion of the kappa gamma matrix,
\begin{equation} \Gamma_\kappa=\Gamma_\ast+\frac{1}{\w}\Gamma_\bullet+
{\cal O}(\omega ^{-2}),
\end{equation}
where $\Gamma_\ast$ was introduced before in \bref{gammaIIA} and, for
type IIA,
\begin{eqnarray} \Gamma_\bullet= -\hga^k
      \Gamma_{a} u^{a}_{k}\Gamma_{\ast}-\frac{1}{2}\cF^{(1)}_{j
k}\Gamma_{11}\hga^j\hga^k\Gamma_{\ast}.
\end{eqnarray}

The symmetries of the non-relativistic lagrangian are a consequence of
the symmetries of the parent relativistic theory and the fact that the
divergent term of the non-relativistic expansion,
$\cL_{\mathrm{GS}}^{\mathrm{div}}=\cL_{\mathrm{NG}}^{\mathrm{div}}-
\cL_{\mathrm{WZ}}^{\mathrm{div}}$,
is a total derivative or is absent when we introduce the coupling to the
RR $C_{p+1}$ form (\ref{cLCp1}).

The supersymmetry transformations of the NR D$p$-brane action for type
IIA \bref{NRNGaction} are given by
\begin{eqnarray}\nn\label{susynr}\nn \delta_\ep\t_-&=&\ep_-,
\qquad\qquad\qquad\delta_\ep\t_+=\ep_+, \\\nn \delta_\ep
x^\mu&=&i\bar{\t}_-\Gamma^\mu\ep_-,\qquad\quad\delta_\ep
X^a=i\bar{\t}_-\Gamma^a\ep_++i\bar{\t}_+\Gamma^a\ep_-,\qquad\delta_\ep
x^a=2i\bt_-\Gamma^a\ep_+, \\
\nn\delta_\ep W&=& -i( \bar{\ep}_+\Gamma_\mu \Gamma_{11}\t_-
+\bar{\ep}_-\Gamma_\mu\Gamma_{11}\t_+)\rmd x^\mu-i
\bar{\ep}_-\Gamma_a\Gamma_{11}\t_-\rmd X^a\\\nn
&&+\frac{1}{6}\bigg[(\bar{\ep}_+\Gamma_\mu\Gamma_{11}\t_-+\bar{\ep}_-
\Gamma_\mu\Gamma_{11}\t_+)\bt_-\Gamma^\mu
\rmd\theta_-\\&&+\bar{\ep}_-\Gamma_a\Gamma_{11}\t_-\left(\bt_-\Gamma^m
\rmd\theta_++\bt_+\Gamma^m \rmd\theta_-\right)\nonumber\\
&&+\bar{\ep}_-\Gamma_\mu\t_-(\bt_-\Gamma^\mu\Gamma_{11}
\rmd\theta_++\bt_+\Gamma^\mu\Gamma_{11}\rmd\theta_-)\nn\\&&+(\bar{\ep}_-\Gamma_a\t_++\bar{\ep}_+\Gamma_a\t_-)\bt_-\Gamma^m\Gamma_{11}\rmd\theta_-\bigg].
\end{eqnarray}
The action \bref{NRNGaction} has also the NR kappa symmetry
\begin{eqnarray}\label{nrkappatrans}
\delta_\kappa\bt_-&=&\bar{\kappa}_-,\qquad\qquad
\delta_\kappa\bt_+=(-)^n\frac12\bar{\kappa}_-\Gamma_\bullet,
\\\nn
\delta_\kappa x^\mu&=&-i\bt_-\Gamma^\mu \kappa_-,
\nonumber\\
\delta_\kappa X^a&=&-i\bt_+\Gamma^a \kappa_-+(-)^ni\bar{\kappa}_-
\frac{\Gamma_\bullet}{2}\Gamma^a\t_-,\qquad\delta_\kappa
x^a=-2i\bt_+\Gamma^a\kappa_- ,\nonumber\\
\delta_\kappa
W&=&i(\delta_\kappa\bt_+\Gamma_\mu\Gamma_{11}\t_-+\delta_\kappa\bt_-\Gamma_\mu\Gamma_{11}\t_+)\hat{e}^\mu
+i\delta_\kappa\bt_-\Gamma^a\Gamma_{11}\t_-u^a \nn\\&&+
\frac{1}{2}(\delta_\kappa\bt_+\Gamma_\mu\Gamma_{11}\t_-+\delta_\kappa\bt_-\Gamma_\mu\Gamma_{11}\t_+)\bt_-\Gamma^\mu
\rmd\theta_-\nonumber\\
&&+\frac{1}{2}\delta_\kappa\bt_-\Gamma_a\Gamma_{11}\t_-(\bt_+\Gamma^a
\rmd\theta_-+\bt_-\Gamma^a \rmd\theta_+)
\nn\\
&&-\frac{1}{2}\delta_\kappa\bt_-\Gamma_\mu\t_-(\bt_-\Gamma^\mu\Gamma_{11}
\rmd\theta_++\bt_+\Gamma^\mu\Gamma_{11} \rmd\theta_-)\nonumber\\
&&-\frac{1}{2}(\delta_\kappa\bt_-\Gamma_a\t_++\delta_\kappa\bt_+\Gamma_a\t_-)\bt_-\Gamma^a\Gamma_{11}
\rmd\theta_-\nn.
\end{eqnarray}
{}From (\ref{nrkappatrans}) we see that $\theta_-$ is a gauge degree of
freedom that can be eliminated by choosing $\theta_-=0$. In this gauge we
can explicitly integrate the WZ term. 
The action for a non-relativistic D$p$ brane in diffeomorphism-invariant
form becomes
\begin{eqnarray}
  S_{\mathrm{NR}}&=&-T_{\mathrm{NR}}\int \rmd \sigma^{p+1}
  \left[\frac1{2}\sqrt{-\det{g}}g^{ij}\pa_i\vec{
X}\pa_j\vec{X}+2i\sqrt{-\det{g}}\bt_+\gamma^i\pa_i\t_+\right.\nonumber\\
&&\left.\qquad\qquad\qquad +\frac1{4}\sqrt{-\det{g}}f_{ij}f_{k\ell
}g^{ik}g^{j\ell }\right].
\end{eqnarray}
where $g^{ij}$ is the inverse of the induced metric
${g}_{ij}=\eta_{\mu\nu}{e}^\mu_j{e}^\nu_j$. This lagrangian is
interacting since the longitudinal scalars $x^\mu(\sigma)$ are coupled to
the transverse scalars $X^a(\sigma)$ and the dynamical fermions
$\theta_+(\sigma)$ via the induced metric $g_{ij}$. The gamma matrices
$\gamma _i$ are the pullbacks of the gamma matrices in spacetime, $\gamma
_i=e_i^\mu \Gamma _\mu $.

In the static gauge  $(x^\mu=\sigma^\mu)$, this theory becomes a
supersymmetric free theory in a flat spacetime of scalars, fermions and
gauge fields.
\begin{equation}
  S_{\mathrm{NR}}=-T_{\mathrm{NR}}\int \rmd
^{p+1}\sigma\left[\frac{1}{2}\eta^{ij}\pa_i\vec{X}\pa_j\vec{X}+2i\bt_+\Gamma^i
\pa_i\t_+ +\frac{1}{4}f_{ij}f_{k\ell }\eta^{ik}\eta^{j\ell }\right].
\end{equation}

Once kappa symmetry is fixed, sixteen of the supersymmetries are linearly
realized while the other sixteen are non-linearly realized. The
non-linear realized supersymmetries are generated by $\epsilon_+$, while
the linearly realized supersymmetries are induced by $\epsilon_-$. The
transformations are
\begin{eqnarray}
 \d\bt_+&=&\bar{\ep}_+-\frac{1}{2}\bar \epsilon _-
 \left( \Gamma ^k \partial_k
X^a\Gamma _a+\frac{1}{2}f_{jk}\Gamma _{11}\Gamma ^{jk}\right),  \nonumber\\
\d X^a&=&2i\bt_+\Gamma^a\ep_-,\nn\\
\d W_i&=&-2i\bar{\ep}_-\Gamma_i\Gamma_{11}\t_+. \label{restransfIIA}
\end{eqnarray}

For type IIB D$p$ branes we obtain the same expressions as for IIA but
with the substitution of $\Gamma_{11}$ by $\tau_3$, at this point we
should note that the substitution must be done before any commutation of
$\Gamma_{11}$ with any other $\Gamma$. The only exception is the kappa
symmetry transformation for the spinor $\t_+$ \bref{nrkappatrans}, which
is written for the IIB case as
\begin{equation}
  \d_{\kappa}\bt_+=\frac12\bar{\kappa}_-{\Gamma_{\bullet}}.
\end{equation}
Consequently, the residual transformation is
\begin{eqnarray}
 \d\bt_+&=&\bar{\ep}_+-\frac{1}{2}\bar \epsilon _-
 \left( \Gamma ^k \partial_k
X^a\Gamma _a+\frac{1}{2}f_{jk}\tau _3\Gamma ^{jk}\right),\nonumber\\
\d X^a&=&2i\bt_+\Gamma^a\ep_-,\nn\\
\d W_i&=&-2i\bar{\ep}_-\Gamma_i\tau_3\t_+. \label{restransfIIB}
\end{eqnarray}

The linearly realized supersymmetries represent the transformations of a
vector multiplet with 16 real supersymmetries in $p+1$ dimensions. The
formulae (\ref{restransfIIA}) and (\ref{restransfIIB}) give a uniform way
for writing these vector multiplet transformations in any dimension using
$D=10$ notation.

\subsection{D1 string}  \label{ss:D1 string}

In this section we will consider explicitly the case of D1 strings. This
case is interesting because we can write explicitly the non-relativistic
Wess-Zumino term, since we can easily integrate the $h_3$ form given by
\bref{IIABhform}, and therefore we can write explicitly the
kappa-symmetric form of the non-relativistic D1 string action.

For the D-string we have $\Gamma_{*}=\Gamma_0\Gamma_1\tau_1$. As in the
general case, we obtain a divergent term for the WZ part, which we can
write explicitly as
\begin{equation}
\cL_{WZ}^{\mathrm{div}}=\varepsilon^{jk}\varepsilon_{\mu\nu}i\bt_-\Gamma^{\nu}\pa_j\t_-\left(\pa_kx^{\mu}+\frac{i}2\bt_-\Gamma^{\mu}\pa_k\t_-\right).
\end{equation}
The $\w^2$ terms of $\cL_{\mathrm{GS}}$ (remember that %
$\cL_{\mathrm{GS}}=\cL_{\mathrm{NG}}-\cL_{\mathrm{WZ}}$) give
\begin{equation}
\cL^{\mathrm{div}}_{\mathrm{GS}}=-\frac12\varepsilon^{jk}\varepsilon_{\mu\nu}
\pa_jx^{\mu}\pa_kx^{\nu}.
\end{equation}
This divergent term can be cancelled by turning on a closed RR $C_2$ form
(\ref{cLCp1}).

The finite part of the kappa-symmetric form of the action is
\begin{eqnarray}
 \label{nrd1}\nn S_{\mathrm{NR}}&=&-T_{\mathrm{NR}}\int\rmd ^2\sigma\bigg[
2\hat{e}i\bt_+\hat{\gamma}^k\pa_k\t_++\frac{1}{2}\hat{e}\hat{g}^{lk}\vu_l\vu_k
+\qquad\\
&&-\frac{1}{4}\hat{e}\hat{g}^{kl}\cF_{lj}^{(1)}\hat{g}^{ji}\cF_{ik}^{(1)}
+2i\varepsilon^{jk}\bt_+\Gamma_a\tau_1\pa_j\t_-(u^a_k-i\bt_+\Gamma^a\pa_k\t_-)
 \bigg].
\end{eqnarray}

If we choose $\t_-=0$ and the static gauge, the action becomes
\begin{eqnarray} \label{d1st}
S_{\mathrm{NR}}=-T_{\mathrm{NR}}\int\rmd
^2\sigma\bigg[\frac{1}{2}\eta^{ij}\pa_i\vec{
X}\pa_j\vec{X}+2i\bt_+\Gamma^i\pa_i\t_+ +\frac{1}{4}f_{ij}f_{k\ell
}\eta^{ik}\eta^{j\ell }\bigg].
\end{eqnarray}
The residual supersymmetry transformation is given by
(\ref{restransfIIB}) for $p=1$.

\section{Conclusions}

Non-relativistic superstrings and D$p$ branes describe a consistent and
soluble sector of the full relativistic string theory. In this paper, we
derived the world-volume theory of non-relativistic supersymmetric D$p$
branes in flat spacetime. This is achieved by considering a suitable
non-relativistic limit of relativistic wound D$p$ branes. The branes are
charged with respect to the $C_{p+1}$ RR form, and we fine-tune the
coupling in such a way that the tension of the D$p$-brane in cancelled by
the RR coupling. This is the cancellation of the superficially divergent
terms in the action. It is important to notice that kappa symmetry is
crucial for this cancellation. The balance between the NG part and the WZ
part of the action needed for kappa symmetry is the same balance that is
necessary for combining these superficially divergent terms in a total
derivative. Then this total derivative can be cancelled by a closed
$C_{p+1}$ form.

Once all the gauge symmetries of the non-relativistic D$p$-brane action
are fixed, the world-volume theory reduces to a supersymmetric field
theory of bosons, fermions and gauge fields in flat spacetime.

The non-relativistic string theory provides a new soluble sector of
string theory where one could test the gauge/gravity correspondence. See
\cite{Gomis:2005pg} for a concrete proposal in the case of $AdS_5\times
S^5$. More in general, it could be interesting to study the
non-relativistic sector of AdS branes, e.g. \cite{Skenderis:2002vf}.

\section*{Acknowledgments.}

\noindent We are grateful to Jaume Gomis, Kiyoshi Kamimura and Paul
Townsend for interesting and very useful discussions.

This work is supported in part by the European Community's Human
Potential Programme under contract MRTN-CT-2004-005104 `Constituents,
fundamental forces and symmetries of the universe'. The work is supported
in part by the FWO - Vlaanderen, project G.0235.05 and by the Federal
Office for Scientific, Technical and Cultural Affairs through the
"Interuniversity Attraction Poles Programme -- Belgian Science Policy"
P5/27, by the Spanish grant MYCY FPA 2004-04582-C02-01, and by the
Catalan grant CIRIT GC 2001, SGR-00065.

Joaquim Gomis acknowledges the Francqui Foundation for the
Interuniversity International Francqui chair in Belgium awarded to him,
as well the warm hospitality at the University of Leuven.

\appendix

\section{Notation and some useful formulae}\label{app}

Here we summarize our notation. Indices are
\begin{eqnarray}
{\rm target\; space}&:& m,n=0,...,9\nn\\
{\rm target\; space, longitudinal}&:& \mu,\nu=0,...,p,\nn\\
{\rm target\; space, transverse}&:& a,b=p+1,...,9\nn\\
{\rm world-volume}&:& i,j=0,...,p.
\end{eqnarray}
The metric in target space and on the world-volume has signature mostly
$+$. The totally antisymmetric Levi-Civita tensor is normalized by
$\varepsilon^{012...p}=+1$, $\varepsilon _{012...p}=-1$.

The $\Gamma^m$ and $\Gamma_{11}$ can be chosen real by taking the charge
conjugation matrix $C=\Gamma_0$, and
\begin{equation}
\Gamma_{11}=\Gamma_{0}\Gamma_{1}...\Gamma_{9}.
\end{equation}
For type IIA theories, $\theta$ is a  Majorana spinor, while for type IIB
theories, there are two Majorana-Weyl spinors $\t_\alpha$ ($\alpha=1,2$)
of the same chirality. The index $\alpha$ is not displayed explicitly.
The Pauli matrices $\tau_1, \tau_2,\tau_3$ act on it. This leads to some
useful symmetry relations as
\begin{equation}
\bar \chi \lambda =\bar \lambda \chi ,\qquad   \lambda =\Gamma _m\epsilon
\ \rightarrow \ \bar \lambda =-\bar \epsilon \Gamma
  _m,\qquad \lambda =\Gamma _{11}\epsilon \ \rightarrow\ \bar \lambda =-\bar \epsilon \Gamma
  _{11}.
 \label{barGamma}
\end{equation}
There are cyclic identities
\begin{eqnarray}
 \sum_{IJK\;\mathrm{cyclic}}\left[\Gamma_m
\t_I\;(\bt_J \Gamma^m\;\t_K)+ \Gamma_m \Gamma_{11}\t_I\;(\bt_J
\Gamma^m\Gamma_{11}\;\t_K)\right]=0, \label{cyclic2}
\end{eqnarray}
and, for type IIB spinors,
\begin{equation} \sum_{I~J~K~\mathrm{cyclic}} \left\{
\Gamma_m\tau_1\t_I\left(\bt_J \Gamma^m \t_K\right) +
\Gamma_m\t_I\left(\bt_J \Gamma^m \tau_1 \t_K\right)\right\} = 0,
\label{CIDB}
 \end{equation}
where $\tau _1$ can also be replaced by $\tau _3$.

We define projections in (\ref{Gammastartheta}), using the matrix $\Gamma
_\ast$ defined in (\ref{gammaIIA}) and in (\ref{gammaIIB}) for type IIA
and IIB, respectively. This matrix squares to $\unity $. Here are some
useful properties:
\begin{eqnarray}
 \Gamma_\ast\t_\pm &=&\pm\t_\pm, \nonumber\\
 \mathrm{type\ IIA} & : & \bar \theta _\pm =\pm
(-)^{\frac{p}{2}+1} \bar \theta _\pm \Gamma _\ast, \qquad
\Gamma _\ast \Gamma _{11}=-\Gamma _{11} \Gamma _\ast,\nonumber\\
&& \Gamma _\ast\Gamma ^\mu =(-)^{\frac{p}{2}+1}\Gamma ^\mu\Gamma
_\ast,\qquad\Gamma _\ast\Gamma ^a =(-)^{\frac{p}{2}}\Gamma ^a\Gamma
_\ast,\nonumber\\
 \mathrm{type\ IIB} & : &\bar \theta _\pm =\mp \bar \theta _\pm \Gamma
_\ast,\qquad \Gamma _\ast\tau _3=-\tau _3\Gamma _\ast,\nonumber\\
&&\Gamma _\ast\Gamma ^\mu =-\Gamma ^\mu\Gamma _\ast,\qquad\Gamma
_\ast\Gamma ^a =\Gamma ^a\Gamma _\ast.
 \label{propGammaast}
\end{eqnarray}
For the D1 string we can also use
\begin{equation}\label{D1b}
\Gamma_{\mu}\tau_{1}\theta_\pm=\pm\varepsilon_{\mu\nu}\Gamma^{\nu}\theta_\pm.
\end{equation}

Differently from \cite{Aganagic:1996nn,Aganagic:1996pe} the differentials
and the spinors have independent gradings. Components of the forms are
defined by
\begin{equation} A_{r}=\frac{1}{r!}A_{i_1\ldots
i_r}\rmd\sigma^{i_1}\ldots \rmd\sigma^{i_r},
\end{equation}
and differentials are taken from the left.

%

\providecommand{\href}[2]{#2}\begingroup\raggedright\endgroup

\end{document}